\documentclass[a4paper,fleqn,usenatbib]{mnras}

\usepackage[T1]{fontenc}
\usepackage{ae,aecompl}

\usepackage{amsmath}
\usepackage{amsfonts}
\usepackage{amssymb}
\usepackage{graphicx}
\usepackage{multirow,bigdelim}
\usepackage{adjustbox}
\usepackage{placeins}
\usepackage{soul}
\usepackage{array}
\usepackage{float}

\newcolumntype{C}[1]{>{\centering\arraybackslash}m{#1}}

\def\CN2{\mbox{$C_N^2 \ $}}

\def\CT2{\mbox{$C_T^2 \ $}}

\def\sigmal2{\mbox{$\sigma ^{2}_{I} \ $}}

\title[High accuracy PWV forecast and sky background]{High accuracy short-term PWV operational forecast at the VLT and perspectives for sky background forecast}

\author[A. Turchi et al.]{
A. Turchi,$^{1}$\thanks{E-mail: alessio.turchi@inaf.it}
E. Masciadri,$^{1}$\thanks{E-mail: elena.masciadri@inaf.it}
P. Pathak$^{2}$\thanks{E-mail: Prashant.Pathak@eso.org}
M. Kasper,$^{2}$\thanks{E-mail: mkasper@eso.org}
\\
$^{1}$INAF - Osservatorio Astrofisico di Arcetri, L.go E. Fermi 5, 50125  Florence, Italy\\
$^{2}$European Southern Observatory, Karl-Schwarzschild-Str.2, 85748 Garching, Germany\\}

\date{Accepted AAAA/BB/CC. Received YYY; in original form ZZZ}

\pubyear{AAAA}

\begin{document}
\label{firstpage}
\pagerange{\pageref{firstpage}--\pageref{lastpage}}
\maketitle

\begin{abstract}
In this paper we present the first results ever obtained by applying the autoregressive (AR) technique to the precipitable water vapour (PWV). The study is performed at the Very Large Telescope. The AR technique has been recently proposed to provide forecasts of atmospheric and astroclimatic parameters at short time scales (up to a few hours) by achieving much better performances with respect to the 'standard forecasts' provided early afternoon for the coming night. The AR method uses the real-time measurements of the parameter of interest to improve the forecasts performed with atmospherical models. We used here measurements provided by LHATPRO, a radiometer measuring continuously the PWV at the VLT. When comparing the AR forecast at 1h to the standard forecast, we observe a gain factor of $\sim$ 8 (i.e. $\sim$ 800 per cent) in terms of forecast accuracy. In the PWV $\le$ 1 mm range, which is extremely critical for infrared astronomical applications, the RMSE of the predictions is of the order of just a few hundredth of millimetres (0.04~mm). We proved therefore that the AR technique provides an important benefit to VLT science operations for all the instruments sensitive to the PWV. Besides, we show how such an ability in predicting the PWV can be useful also to predict the sky background in the infrared range (extremely appealing for METIS). We quantify such an ability by applying this method to the NEAR project (New Earth in the Alpha Cen region) supported by ESO and Breakthrough Initiatives.
\end{abstract}

\begin{keywords} site testing -- atmospheric effects -- methods: data analysis -- methods: numerical 
\end{keywords}



\section{Introduction}
\label{intro}

Quantification of the water vapour is particularly critical for ground-based observations in the infrared spectrum (IR, sub-millimetric and millimetric ranges), since it is one of the atmospheric constituents that mostly affects the atmosphere transmission at those wavelengths. The transmission/opacity of the atmosphere is mainly caused by the absorption of radiation from H$_{2}$O. On the other side, the water vapour is also one of the causes that contributes to the sky background, due to the re-emission of the absorbed energy which impacts on the sensitivity of IR observations, and the strength of the process is dependant on the H$_{2}$O abundance in the atmosphere.
The vertical distribution of the water vapour in the atmosphere can be described by the vertical stratification of the mixing ratio M or the absolute humidity AH. For most of the practical astronomical applications, the relevant parameter is the integral of these quantities along the whole atmosphere, the so called precipitable water vapour (PWV). The forecast of this parameter can be therefore extremely useful to optimise the potentialities and the efficiency of instruments running at these wavelengths. The PWV can be forecasted to support science operations by using atmospherical models, and several studies employing mesoscale models to reconstruct the PWV can be found in the literature (\citet{cucurull2000}, \citet{gonzalez2013}, \citet{giordano2013}, \citet{perez2015,perez2018}, \citet{pozo2016}).

In a recent paper \citep{turchi2019}, the authors used the non-hydrostatical atmospherical model Meso-Nh to forecast the PWV above Cerro Paranal (Chile), site of the Very Large Telescope (VLT), on a time scale of a 6 to 15 hours in the future. More precisely, the forecast of the PWV for the coming night is provided early in the afternoon of the day before. A statistical analysis has been performed above Cerro Paranal, site of the VLT, on a sample of 240 nights uniformly distributed in different years (2013 and 2017) and statistical operators (bias, RMSE and $\sigma$) have been quantified in different ranges: all PWV values, PWV $\le$ 5 mm and PWV $\le$ 1 mm.
The results of the previous analysis revealed that the proposed model and the approach were extremely promising. The smaller is the threshold of the analysed interval, the smaller is the RMSE achieved, up to a value of RMSE = 0.27 mm in the most challenging region, i.e. when PWV $\le$ 1 mm. This indicates that the proposed approach seems very efficient to reconstruct the most interesting values in the astronomical context, i.e. the extremely low values of the PWV. In the same paper, we proved that the Meso-Nh model provides a substantial improvement in performances (roughly a factor 2 i.e. 100 per cent) with respect to the Atmospheric General Forecast Model of the European Centre for Medium Range Weather Forecasts (ECMWF).

More recently, a method has been proposed \citep{masciadri2020} aiming to improve the model performances in forecasting some relevant atmospheric and astroclimatic parameters on short time scales (order of a few hours in the future, typically 1h or 2h) using an autoregressive (AR) technique that simultaneously uses the forecast obtained with an atmospheric model on longer time scales\footnote{Typically forecasts available early afternoon for the next night} and the real-time measurement of the specific forecasted parameter. Such a method has been successfully tested on the Large Binocular Telescope (LBT) and it has been implemented in the ALTA Center, the operational forecast system conceived to support science operations of this telescope. The time scale of one hour in the future is the most critical for the queue modes of top-class telescopes, because this is the time scale on which decisions are typically taken, either to switch from a program to another or from an instrument to another. 

Results of that study indicate that important gains in the forecast reliability can be reached using the autoregressive approach, even if the gain (between 2.70 and 4.90 in terms of RMSE reduction) depends on the specific atmospheric parameter. However, the LBT site has no real-time measurements of the PWV. We still do not know, therefore, what might be the gain obtainable in the prediction of the PWV using this technique. The PVW is nightly monitored at the VLT with a Low Humidity and temperature Profiling microwave radiometer (LHATPRO). This instrument has been validated by \citet{kerber2012} and it has been engineered for monitoring dry sites.\\

In this paper we evaluated the impact of this method on the PWV at the VLT by quantifying the gain obtained on the prediction accuracy of this parameter with respect to the standard forecasts (forecast at long time scales) and also with respect to the persistence method. The persistence method assumes that the value of a specific parameter remains equal to the value measured at the present time t$_{0}$ for the successive hour. Increasing the forecast accuracy, even if over short time windows, can definitely help the flexible scheduling and the identification of the best PWV windows to optimise infrared observations that critically depend on this parameter.

Besides, we explore further contexts in which the forecast of the PWV might be critical and crucial. We focused therefore our attention on the project NEAR \citep{kasper2017}, a challenging experiment promoted by ESO and Breakthrough Initiatives\footnote{\href{http://breakthroughinitiatives.org}{http://breakthroughinitiatives.org}} that aims to search for exoplanets in the habitable zone (HZ) around $\alpha$ Centauri in N band. Earth-type planets, with characteristics similar to our own planet, typically have orbits within [0.1-1] au of their star (red dwarf or solar type). Such kind of planets can not be generally detected from 8-10 m class telescopes, as we need a contrast of 10$^{-5}$ - 10$^{-7}$ in N band for stars within 10 pc from Earth (10$^{-5}$ for red dwarfs and 10$^{-7}$ for solar-type stars), but they represent an important science case for the ELT. In the case of the $\alpha$ Centauri star system ($\alpha$ Centauri A and B), however, because of its proximity (1.35 pc) to Earth, the angular separation of a potential planet should be of the order of 1\arcsec and the observation of a planet in the HZ should in principle be possible also with a 8 m class telescopes by pushing to the extreme the possibile contrast and sensitivity attainable on such tools. The concept of NEAR project implies the modification of VISIR instrument, that consists on combining the instrument with the adaptive optics using the deformable secondary of UT4, adding a phase mask coronograph optimised for N band and including a new chopper system for noise filtering \citep{kasper2017}. In the analysis presented in this contribution, we used all the NEAR project observations related to the commissioning run on the 23 May - 10 June 2019 period plus the nigh of the 26 June 2019, for a total of 18 nights, and all the data-set related to the run in shared risk basis to the community for the Science Verification (SV) related to the 13-19 December 2019 run, for a total of 7 nights. 

The wavelength window has been selected between 10 $\mu$m and 12.5 $\mu$m, where the SNR is supposed to be maximum. Preliminary observations done during the commissioning seemed to indicate that the PWV has an impact on the sky background even if in principle we are in a region dominated by the telescope emission. The idea is therefore to analyse the sky background and the PWV and to establish if we can retrieve any correlation or relation between the two quantities. This might provide new insights on the PWV role for the instrumentation for the new generation class telescopes and eventually help us in identifying other applications of the PWV forecasts.

The paper is organised as follows. In Section \ref{sec:mod_conf} we briefly present the configuration of the numerical model used in the present study. In Section \ref{sec:obs} we discuss the measurements provided by VLT instrumentation. In Section \ref{sec:ar} we briefly summarize the principle of autoregression technique that we used to enhance PWV forecast over a timescale of 1h. In Section \ref{sec:res} we describe the model performances obtained using the autoregression technique and we compare results obtained with the standard forecast approach and with method by persistence. In Section \ref{sec:NEAR} we quantify the correlation between the sky background measured with NEAR during the commissioning and the PWV and we discuss the importance of the PWV forecast in this context. Finally, in Section \ref{concl} we present our conclusions.\\

\section{Model Configuration}
\label{sec:mod_conf} 

The atmospheric model used to forecast the atmosphere is Meso-NH\footnote{\url{http://mesonh.aero.obs-mip.fr/mesonh/}} (MNH) \citep{lafore98,lac2018}. It is a non-hydrostatic mesoscale model that simulates the time evolution of a finite 3D atmospheric volume over a specific geographical area. The specific model configuration was basically the one reported in \citep{turchi2019}, thus we report here just the most relevant elements. We performed the simulations over Cerro Paranal (VLT site) located at coordinates (24.62528 S, 70.40222 W) at an height of 2635~m above sea level.\\

We use a grid-nesting technique \citep{stein2000} that consists of using a set of different imbricated domains, described in Tables \ref{tab:resol}, with a digital elevation model (DEM, i.e. orography) extended on smaller and smaller surfaces having a progressively higher horizontal resolution. In this way, using the same vertical grid resolution, we achieve the highest horizontal resolution on the innermost domain extended on a limited surface around the summit to provide the best possible prediction. Each domain is centered on the telescope coordinates.\\

\begin{table}
\caption{Horizontal resolution of each Meso-NH imbricated domain at Cerro Paranal (VLT).} 
\label{tab:resol}
\begin{center}       
\begin{tabular}{cccc} 
\hline
\rule[-1ex]{0pt}{3.5ex}  Domain & $\Delta$X (km) & Grid points & Domain size (km) \\
\hline
\rule[-1ex]{0pt}{3.5ex}  Domain 1 & 10 & 80x80 & 800x800 \\
\rule[-1ex]{0pt}{3.5ex}  Domain 2 & 2.5 & 64x64 & 160x160 \\
\rule[-1ex]{0pt}{3.5ex}  Domain 3 & 0.5 & 150x100 & 75x50 \\
\hline
\end{tabular}
\end{center}
\end{table}

The DEM used for domains 1 and 2, is the GTOPO\footnote{\url{https://lta.cr.usgs.gov/GTOPO30}}, with an intrinsic resolution of 1~km. In domain 3 we use a digital elevation model with an intrinsic resolution of 500~m (16\arcsec). \\
We use the the grid-nesting 2-way option that allows a mutual interaction between each couple of father and son domains i.e. each couple of contiguous domains. The 2-way configuration is known to provide more consistent and realistic results than the 1-way configuration \citep{stein2000,harris2000}.\\
We use 62 vertical levels on each domain, starting from 5~m above ground level (a.g.l.). The levels have a logarithmic stretching of 20\% up to 3.5~km a.g.l. From this point onward the model uses an almost constant vertical grid size of $\sim$ 600~m up to $\sim$23~km, which is the top level of our domain. The grid mesh deforms uniformly to adapt to the orography, so the actual size of the vertical levels can stretch in order to accommodate for the different ground level at each horizontal grid point.\\

In this study we implemented the \cite{turchi2019} model set-up in an automatic forecast system in order to run consecutively over a whole year. The system takes care of all the phases of the forecast, from the model initialisation, the data assimilation and grid configuration, the parallel simulation and the final post-processing phase and the production of the final outputs. The system is designed in order to guarantee the delivery of the forecasts early in the afternoon before the start of the observing night. The forecasts computed in this way will be referred as "standard configuration" in the following sections and are the benchmark over which we will evaluate the performance of the AR method \cite{masciadri2020}.\\
The model is initialised with daily operational forecasts computed by ECMWF general atmospheric forecast model extended on the whole globe. Simulations cover the night time at VLT, which is relevant for astronomical observations on each site. The date of each simulation in this paper is identified by the UT day "J'' in which the night starts. For the Cerro Paranal case, in this study we simulate 18 hours initializing the model at 18:00 UT of the day "J'' with ECMWF forecasts computed at 12:00 of the same day. We force the model each consecutive 6 hours with data coming from the ECMWF and we treat/analyse results in the interval [00:00 - 09:00] UT of day "J+1''. This interval permits to fit the nighttime during the whole solar year.\\

Meso-NH computes the PWV using the vertical profiles of water vapor mixing ratio M (kg/kg), pressure P (Pascal) and temperature T (Kelvin degrees), using the following equation \ref{eq1}.
\begin{equation}
  PWV=-\frac{1}{g\rho_{H_2O}}\int_{P_0}^{P_{top}}M dP
  \label{eq1}
\end{equation}
PWV is expressed in mm. In the above equation $\rho_{H_2O}=10^3 kg/m^3$ is the water density and $g=9.81$ m$\mathbf s^{-2}$ is the standard gravity acceleration. We integrate between the ground level pressure $P_0$ and the top level ($\sim$20~km a.g.l) pressure $P_{top}$. We note that the water vapour scale height is in the range 1.5-2.5~km. Above the latter height the water content decreases drastically and is typically negligible above 10~km \citep{Querel2016}.\\
The PWV value is provided with a time sampling of two minutes of simulated time and it is calculated in the innermost domain having a horizontal resolution of 500~m. Following results obtained by \citep{turchi2019} - Eq.2, we applied the post-processing correction obtained in statistical terms aiming to remove systematic errors of the forecast.


\section{Measurements and Instrumentation}
\label{sec:obs}

The instrument used as a reference in this paper is the Low Humidity and Temperature Profiling microwave radiometer (LHATPRO) that has been installed at Cerro Paranal in 2011 \citep{kerber2012} and since then runs continuously 24/7 providing measurements stored in the ESO archive\footnote{\url{http://www.eso.org/asm/ui/publicLog?name=Paranal}}. This instrument is completely automated and is manufactured by Radiometer Physics GmbH. It uses multiple microwave channels in the frequency bands of 183 GHz (H$_{2}$O) and 51-58 GHz (O$_{2}$) in order to retrieve, among others, the humidity and temperature profiles up to 10~km of altitude above the ground level. Measurements are taken on 39 vertical levels with a resolution that varies from 10~m at the ground level up to 1~km at the topmost height. For more detailed description see \citet{Rose}. As explained in \citet{kerber2012}, the 183 GHz line is extremely important because it allows to resolve the extremely low levels of PWV present on a dry site such as Paranal (median value is around 2.4~mm).\\
LHATPRO was validated for astronomical use in 2011 \citep{kerber2012} against radiosoundings, showing a good correlation with measurements, an accuracy of 0.1~mm and a precision of 0.03 mm) for the PWV. The instrument decreases its accuracy starting from 15~mm, and starts to saturate for PWV values above 20~mm. These correspond however to very rare bad weather events at Paranal (typically below 15~mm). In this study we considered 15~mm as a top threshold for our analysis.

LHATPRO measures the PWV with a sample rate of 5 s, while Meso-NH samples the PWV every 2 minutes. In order to have a common time frame, which is fundamental in the following AR method implementation, we decided to resample both data-sets over 1 minute timestep, averaging on the fast LHATRO sample rate and interpolating over the Meso-NH slower one.

\section{Autoregressive method}
\label{sec:ar}

One of the goals of this paper is to apply the autoregressive method \citep{masciadri2020} to the PWV at the VLT in order to quantify the gain in forecast accuracy over a short time window, which is fundamental to help the optimal planning of observations critically dependent on atmospheric water vapour content in flexible scheduling. The details of the implemented method are reported in \citep{masciadri2020} and here we report a short summary. 
We define a function $X_{t}$ that depends on the difference between the standard configuration forecast ($M$) and measurements ($O$) calculated at time {\it t}:
\begin{equation}
  X_{t}=(O_{t}-M_{t}).
\end{equation}
The AR method allows to predict the future value of the above function through a polynomial equation with $P$ regressor coefficients $a_{i}:$
\begin{equation}
  X_{t+1}=\sum_{i=1}^{P} a_{i}(O_{t-i+1}-M_{t-i+1})
\end{equation}
The $a_{i}$ regressors are computed through a least mean square minimization over a finite number $N$ of past nights, then the AR PWV prediction $A$ is obtained as $X^{*}_{t+1}=M_{t+1}+X_{t+1}$. 

At each time $t$ during the night, all the measurements for $t^\prime \leq t$ and all the standard configuration predictions until of the end of the night are available. In order to perform a prediction over $L$ hours in the future, we advance the time series $X^{*}$ for one hour in the future, then we update again the regressor coefficients adding sequentially the AR predicted values between $t$ and $t+1$ hour to the measurements collected util time $t$. This is done in order to contain the AR divergence from the measurements. Then we repeat the above prediction.\\
We used the same approach described in \cite{masciadri2020} and applied to Mt.Graham, site of the Large Binocular Telescope i.e. the AR is computed each hour with the same number of regressors (50) and using the same number of nights in the past (N=5). As explained in the previous paper, it has been observed that the gain of the AR with respect to the standard forecast, in terms of RMSE reduction, is often grater than 1 even after 4 hours in the future. In this paper, we extended the estimate of the RMSE as a function of $\Delta$T (the time in the future from any given present time) up to 6 hours. We only consider nighttime and, for any date "J'', we do not have standard forecast data before 18 UT; moreover the method can only run successfully if there is at least an hour of measurements that overlaps with the standard prediciton starting from $X_{t}$ backward, so the AR forecast cannot start before 19 UT on each day. If the previous condition is not met, we assume automatically that we do not have enough data to perform the AR prediction.\\
Within the previous constraints, we notice the larger is $\Delta$T, the smaller is the sample on which one can compute the statistics. In other words the sample on which RMSE is computed at $\Delta$T=1h is richer than that on which is computed at $\Delta$T=6h\footnote{Both samples remain in any case statistically significative.}.

To produce the AR forecasts we updated the automatic system described in section \ref{sec:mod_conf} by adding the AR method, again simulating regular operations, using the real-time measurements available at VLT. The aforementioned exercise allowed us to successfully validate the automatic forecast implementation and obtain a large set of simulations over a whole year at Cerro Paranal.\\

\section{Results}
\label{sec:res}
In this section we aim to quantify the performances of the AR method explained in section \ref{sec:ar}. The AR method requires a continuous sample of simulated nights to work efficiently, due to the need of a buffer of past nights over which to compute the autoregressor coefficients. For this reason we considered a total sample size of 365 nights between 2018/08/01 and 2019/07/31 (UT). On 10 nights we couldn't run the AR method due to missing data,
so the final sample size used in the analysis is 355 nights. We performed the analysis in terms of classical statistical operators RMSE, $\sigma$, bias and scatterplots between measurements and forecasts in both configurations (standard and AR) in order to compare the different forecast accuracies of the two methods.\\

In order to be consistent with previous studies done on the PWV at Cerro Paranal \citep{turchi2019} and the study carried out with the AR method \citep{masciadri2020}, we performed a 20 minutes resampling over both measurements and predictions (standard configuration and AR) before computing the statistical operators and the scatterplots. The main reason behind this is to filter out the high frequencies and put in evidence the forecast trend, which is the most relevant information from the user point of view.\\

Considering that the median value of PWV over Cerro Paranal is 2.5~mm, as reported in figure \ref{fig:cumdist}, we divided the sample into three ranges: the first takes into consideration all PWV values below 15~mm, which accounts for almost all the PWV values registered over Cerro Paranal and, as explained in section \ref{sec:obs}, is the value range in which the instrument is most accurate; the second range takes into account all PWV values below 5~mm, which includes the most common PWV values; the third range considers all PWV values below 1~mm, which are to be considered rare extremely low PWV events in which water is almost absent from the vertical column over the telescope, and can effectively allow observations practically unobstructed by water absorption. In the sample analysed in this study we calculated that this corresponds to 8 per cent of the total PWV measured values. In Annex A, we repeated the analysis on the last five years and we verified  that the percentage of cases in which the PWV $\leq$ 1mm is of order 5 - 10 per cent, therefore the statistics of the sample analysed in this study is reliable. The relative rarity of these extremely low PWV clearly shows the benefit of an accurate model forecast in order to take advantage of low PWV time windows as soon as they appear. The statistical analysis was performed over all the three different ranges.\\

\begin{table}
 \begin{center}
 \caption{In each column are the statistical indicators computed over the 355 nights sample considered in this study. The indicators are computed  on a sample filtered by selecting LHATPRO measurements with PWV$\leq$15~mm sample , on a sample with PWV$\leq$ 5~mm and a sample with PWV$\leq$ 1~mm. We present values computed first on the standard configuration forecast, then on the enhanced AR forecast extended on 1 hour in the future. In the latter case, in parentesis, we report the gain factor with respect to the standard configuration on the specific statistical indicator.
 } 
 \begin{tabular}{c|ccc}
 \hline
 \multicolumn{1}{c}{} & RMSE (mm) & BIAS (mm) & $\sigma$ (mm) \\
 \hline
 \multicolumn{4}{c}{\bf Standard configuration} \\
 \hline
   All PWV & 1.02 & -0.25 & 0.99 \\
   PWV$\leq$ 5~mm &  0.67 & -0.08 & 0.66 \\
   PWV$\leq$ 1~mm &  0.33 & 0.05 & 0.33 \\
 \hline
 \multicolumn{4}{c}{\bf AR 1h-forecast} \\
 \hline
   All PWV & 0.13 (7.85) & -0.01 (24.0) & 0.13 (7.62)\\
   PWV$\leq$ 5~mm &  0.08 (8.38) & -0.01 (8.0) & 0.08 (8.25)\\
   PWV$\leq$ 1~mm &  0.04 (8.25) & 0.00 (NA) & 0.04 (8.25)\\
 \hline
 \end{tabular}
 \label{tab:rawstat}
 \end{center}
\end{table}

\begin{figure*}
\centering
\includegraphics[width=.9\textwidth]{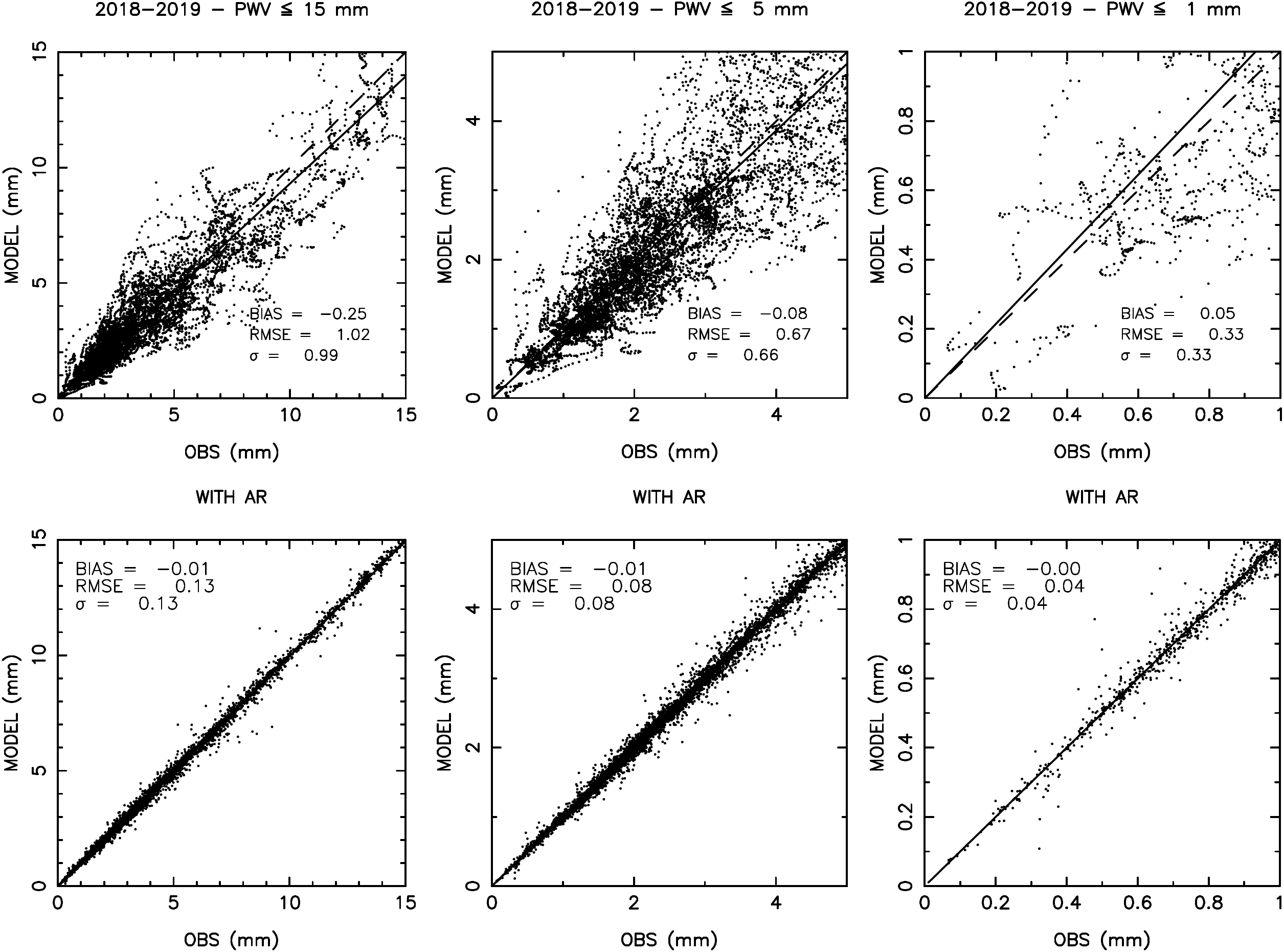}
\caption{First row: scatterplots computed between forecasts and observations on the whole sample of nights (355). In the first row the forecasts in the standard configuration, in the second row the results obtained with the AR method on a time scale of 1 hour. The statistical operators are calculated in three different ranges: for PWV $\leq$ 15~mm (left), for PWV $\leq$ 5~mm (centre) and for PWV $\leq$ 1~mm (right). The straight line corresponds to the regression line computed on the data points.}
\label{fig:scatter}
\end{figure*}

In Fig. \ref{fig:scatter}, first row, we show the scatterplots obtained comparing MNH outputs (standard configuration), which serve as a reference for this study, with measurements on the three PWV ranges. Statistical indicators are also reported in Table \ref{tab:rawstat} - top. The results of the same analysis performed on the AR forecasts extended on 1 hour in the future is reported in Table \ref{tab:rawstat} - bottom, associated to the respective gains with respect to the standard forecast (Fig. \ref{fig:scatter} - bottom) reported in parenthesis. We see, by comparison with the standard configuration results, that the correlation between the AR prediction and measurements in significantly enhanced. The RMSE is in the range [0.13 - 0.04~mm], which is comparable or even below the accuracy of the LHATPRO instrument itself (which \citet{kerber2012} reported to be $\sim$0.1~mm\footnote{Of course the instrument accuracy is valid for all reported PWV values, while the forecast accuracy of 0.04 mm is for PWV $\leq$ 1~mm} and is computed with respect to the measured value.).  
By looking at Table \ref{tab:rawstat} results show that the AR prediction that we implemented provide a considerable gain on a time scale of 1 h with respect to the standard forecast. The maximum gain is 8.38 and it is 8.25 for the PWV $\leq$ 1 mm. The ability to predict the rare extremely low PWV events with negligible discrepancy from the measured value (0.04~mm RMSE) is of extremely helpful in observation scheduling at VLT.\\

\begin{figure*}
\centering
\includegraphics[width=.8\textwidth]{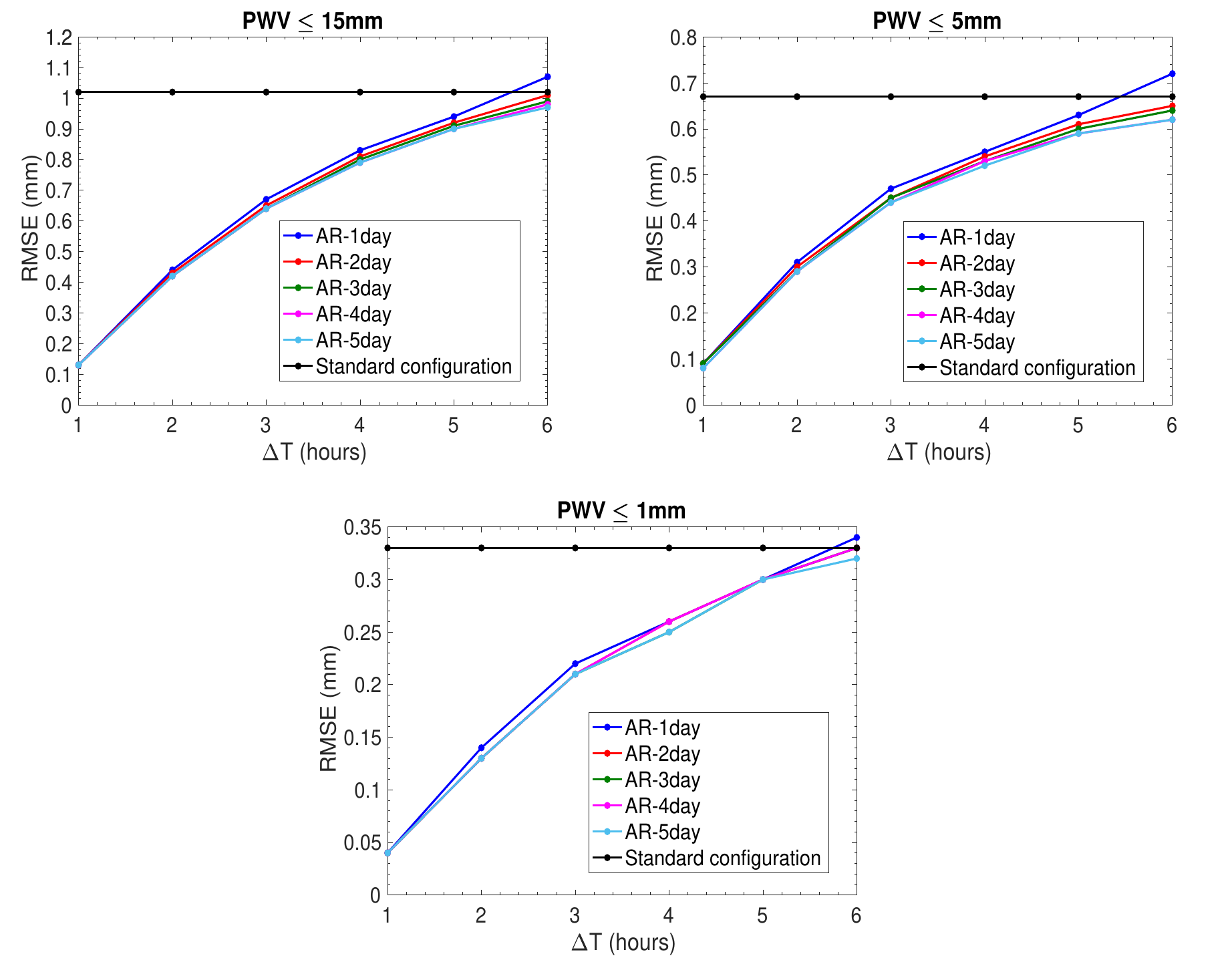}
\caption{RMSE obtained for the AR method with respect to different forecast times ($\Delta$T) from 1 to 6 hours, and with different buffer sizes (different colour curves). The black line is the reference RMSE obtained with MESO-NH standard forecasts. The RMSE is computed over the PWV $\leq$ 15~mm sample (top-left), on the PWV $\leq$ 5~mm sample (top-right) and the PWV $\leq$ 1~mm sample (bottom-centre).}
\label{fig:ARscale}
\end{figure*}

Following the same scheme used in \citep{masciadri2020} in Fig.\ref {fig:ARscale} we report the RMSE obtained by applying the AR method over different forecast time scales, from 1 to 6 hours in the future. The horizontal black line corresponds to the RMSE error obtained with the MNH model forecast from the day before, while the coloured lines correspond to the AR forecast performed with an increasing number of past nights used in the AR coefficients computation (from 1 to 5). As expected, we observe that the AR forecast accuracy tends to decrease with a longer forecast timescale in the future, becoming similar to the mesoscale MNH forecast (standard configuration). The gain factor is still relevant (still a 30\% reduction in the RMSE) up to 3 hours in the future and tends to become negligible after 6 hours. It is interesting to observe that results published in \citep{masciadri2020} relative to other atmospheric parameters (temperature, relative humidity, wind speed and direction), reported a maximum gain of 4.9, while in this study we observe a much higher factor of the gain for the PWV in terms of RMSE reduction (almost double). Of course we can not state, at present time, if this is due to the different site (Mt. Graham versus Cerro Paranal) or to the nature of the parameter itself.

For completeness we remember that the scattering plots related to the standard forecast have been calculated here only to be able to quantify the gain of the AR method.  A dedicated study on the standard forecast of PWV at VLT has been performed by \cite{turchi2019}. We refer the reader to that paper for details. We just observe that results found here are perfectly consistent. The small differences in the statistical operators (bias, RMSE and $\sigma$) are due tot he fact that the samples are not the same. It is important to note that in the case of \cite{turchi2019}, in order to analyse the window PWV $\leq$ 1~mm, the authors enriched the sample size in the low PWV range by including extra nights in which a PWV $\leq$ 1~mm has been observed to get a more significative statistical sample, finding a RMSE = 0.25 mm. We can not do the same in this study as we are forced to select consecutive dates due to the AR method requirements. The enrichment cannot therefore be replicated in the same way.

Also, from figure \ref {fig:ARscale} it is evident that the AR performance increases with the number of past nights considered (N) in the AR coefficient computation even if in a less important way than what observed in the study performed at Mt.Graham for other atmospheric parameters \cite{masciadri2020}. The improvement in performances is absent on a time scale of 1  hour and becomes very small for N $\geq$  3. In applying this method to an operational forecast setup for Cerro Paranal, we can however use N = 5 as the computational cost between 3 and 5 nights buffer is not relevant. As expected the mesoscale forecast tends to be more efficient on longer timescales, and also has the advantage of being available early in the afternoon for the telescope planning. This comparison however clearly explains the need for different forecast schemes depending on the different use cases. For the PWV the AR forecast is advantageous with respect to the standard one on time scales up to around 6 assuming that the night has already started.\\

\begin{figure*}
\centering
\includegraphics[width=.8\textwidth]{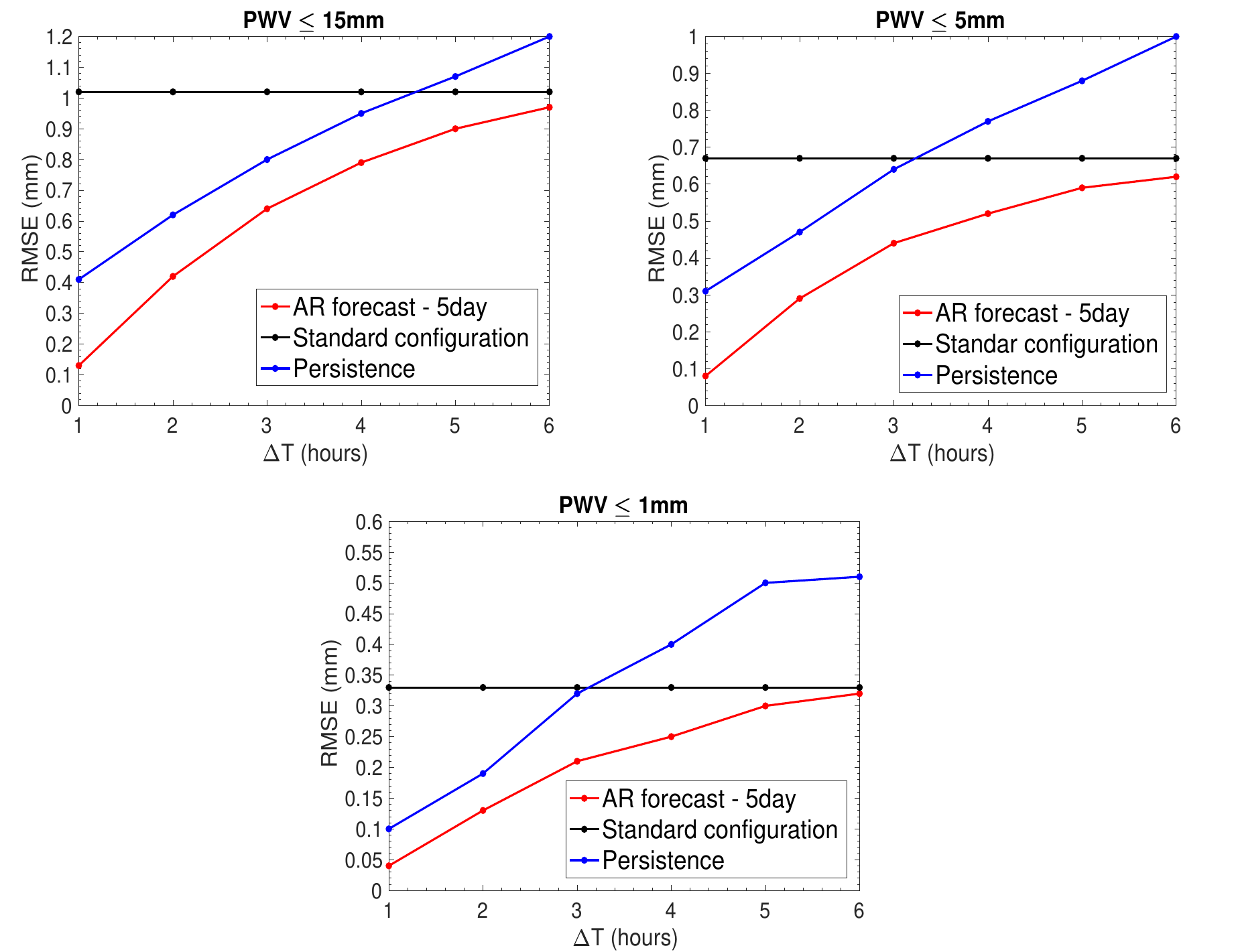}
\caption{RMSE obtained for the persistence method (blue line) and the AR method (red line) for different forecast times $\Delta$T from 1 to 6 hours. The black line is the reference RMSE obtained with MESO-NH standard forecast. In the AR method N = 5. The RMSE is computed over the PWV $\leq$ 15~mm sample (top-left), on the PWV $\leq$ 5~mm sample (top-right) and the PWV $\leq$ 1~mm sample (bottom-centre).}
\label{fig:ARpersist}
\end{figure*}

Fig. \ref{fig:ARpersist} shows the comparison of the AR method versus the method by persistence. The latter forecast method assumes that, at each full hour as in the AR case, we can forecast the future values of PWV by considering the present time PWV value as constant, i.e. assuming that PWV will not change in the successive 1 to 6 hours. We observe that the persistence method clearly loses against the AR method in all configurations. Moreover, as in the AR method case, the persistence forecast method decreases in performances with forecast time ($\Delta$T), but it becomes worse that the standard forecast much before than the AR method i.e. only after 3-4 hours. This is particularly evident for the PWV $\leq$ 1~mm case. In the latter case the performance scaling of the persistence method is more irregular than in the other two cases, however this may be due to the fact that the statistics over this particular sample is lower with respect to the other two cases. We conclude therefore that, even if the forecast by persistence introduces some improvements in the forecast performances at short time scale with respect to the standard forecast, the AR method definitively provides much better performances (a factor 3 to 2 better) with respect to the method by persistemce. These results are consistent with what found for other atmospheric parameters above Mt.Graham \citep{masciadri2020}.\\

\begin{figure*}
\centering
\includegraphics[width=1.0\textwidth]{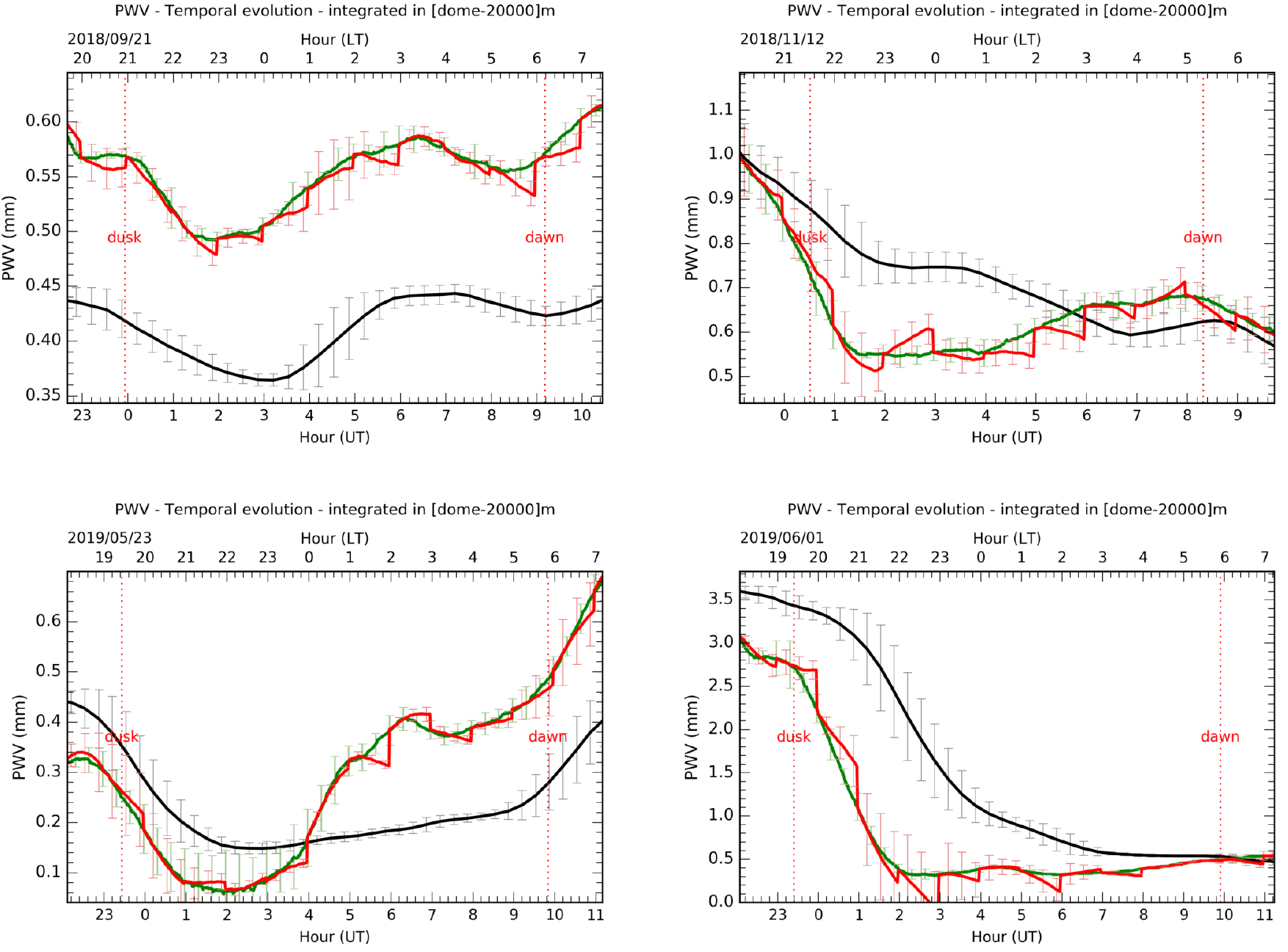}
\caption{Time evolution of PWV during few test nights with low PWV events. Black line shows the standard forecast, green line shows the observations and red line shows the forecast with the AR method on a 1h time scale. The time evolution of the AR forecast is reconstructed by aligned sequentially all the 1-hour forecasts performed at each hour during the night. The error bars represent the variability of the PWV during a 20-minutes interval. The date of each figure corresponds to the start of the night at VLT. The top x-axis is time in LT, while the bottom x-axis is time in UT. The times corresponding to dawn and dusk are shown in each figure through the red dotted vertical lines.}
\label{fig:ARlow}
\end{figure*}

To illustrate the capabilities of the AR forecast in correctly reconstructing the short-time PWV predictions we report here some practical examples. Fig. \ref{fig:ARlow} shows a few examples associated to extremely low PWV events. The time evolution of the AR forecast is reconstructed by joining sequentially all the 1-hour forecasts performed at each hour during the night. In this pictures, we can appreciate that the standard forecast is already quite close to the observations but the AR method is able to improve the forecast accuracy in a not negligible way. The AR prediction helps in correcting small systematic errors, e.g. 2018/09/21 case, in situations were the standard forecast correctly predicts the PWV evolution trends. In other cases, e.g. 2018/11/12 and 2019/06/01, the AR prediction can anticipate rapid drops of the PWV values, giving a clear advantage in the program switching of the telescope during the observing night. Specifically in night 2019/06/01, the standard prediction is lagging behind $\sim$2 hours with respect to the measurements, and the AR is very efficient in correcting this temporal shift. The most interesting case is the night 2019/05/23, were the AR forecast was able to reconstruct an extremely rare low PWV event (PWV = 0.1 mm) lasting for 4 hours which basically allows for ground observations completely unobstructed by the atmospheric water.\\

\begin{figure*}
\centering
\includegraphics[width=1.0\textwidth]{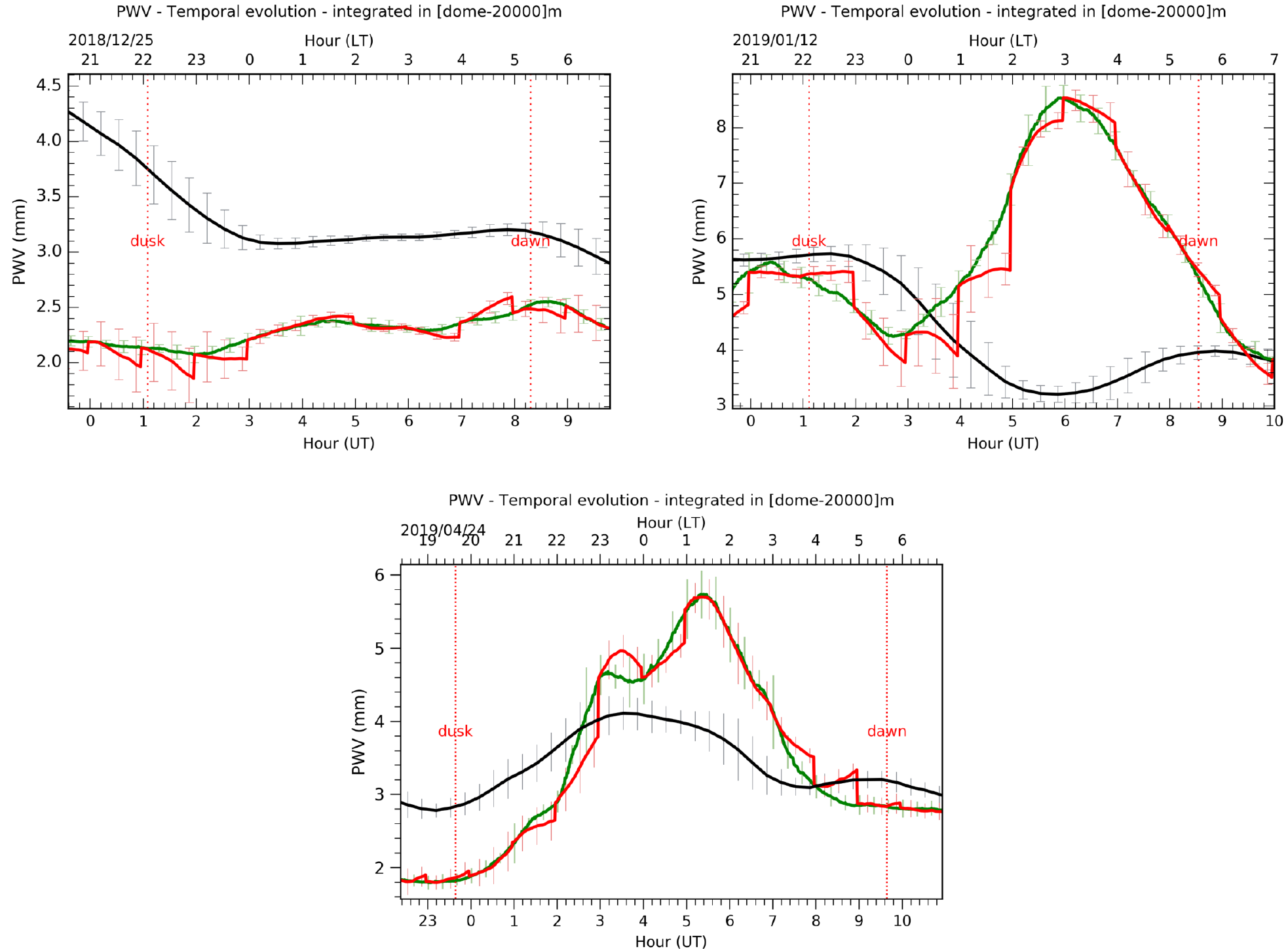}
\caption{As Fig.\ref{fig:ARlow} but the logic of the selection has been that to identify a few test nights were the AR forecast is able to correct discrepancies between the standard forecasts and the observations.}
\label{fig:ARfail}
\end{figure*}

Fig. \ref{fig:ARfail} show on the contrary some examples in which the standard forecasts show a larger deviation from the measured PWV values. In these cases the model may either present large biases with respect to the measurements, or be unable to predict events in which the PWV value unexpectedly changes due to unpredictable weather fluctuations. In these situations the AR forecasts also helps in correcting these large discrepancies and recover an accurate forecast and trend reconstruction in an extremely efficient way. The AR forecast is able indeed to correct the model prediction and reconstruct the evolution of the PWV, even in presence of large fluctuations were PWV varies by 4-5~mm in 2-3 hours (e.g. 2019/01/12). This level of performance allows for an extremely accurate observation scheduling which saves observing time and optimise the throughput of scientific programs critically dependent on PWV.

Results obtained in this study greatly improve our ability in reconstructing the PWV in advance with potential important consequence in terms of the efficiency of the queue mode managing of top-class telescopes such as the VLT. Such a high level of reliability in the PWV forecast can be useful for all the instruments of the VLT running in the IR, sub-millimetric and millimetric ranges. We think therefore about the VLT Imager and Spectrometer for the mid-Infrared (VISIR; \citep{lagage2004}) which operates in the (5-20 $\mu$m) ranges.  At the same time, this can be useful for a set of other instruments running at shorter wavelengths (visible or near-infrared) that benefit from a low value of the PWV, such as CRIRES \citep{kaeufl2004}, X-SHOOTER \citep{vernet2011}, MUSE \citep{henault2003} and VLTI GRAVITY \citep{eisenhauer2008}.\\

\section{NEAR project: sky background}
\label{sec:NEAR}

We present in this section an application of the PWV forecast to the NEAR IR project which has run at VLT, as explained in section \ref{intro}, operating in the 10-12,5~$\mu$m wavelength range. As reported in \cite{kasper2017} a new spectral filter has been implemented in NEAR to maximise the signal-to-noise ratio (SNR). The selection of the bandpass is the result of a compromise. From one side it should be interesting to go towards a wide bandpass to increase the number of photons collected. On the other side, it should be preferable to skip-off some parts of the spectrum. More precisely, below 10 $\mu$m the sky background increases (\cite{kasper2017}- Fig.4) and the absorption from ozone O$_{3}$ diminishes the atmospheric transmission. Above 12.5 $\mu$m the absorption from CO$_{2}$ becomes an important limitation for the observations. In the context of the NEAR project, therefore, the band [10 - 12.5] $\mu$m has been selected. In this band the background is dominated by the telescope emission and the sensitivity to the other atmospheric conditions is weak, thus the PWV value is one of the main elements left that must be accounted for. 
%

\begin{figure}
\includegraphics[width=0.4\textwidth]{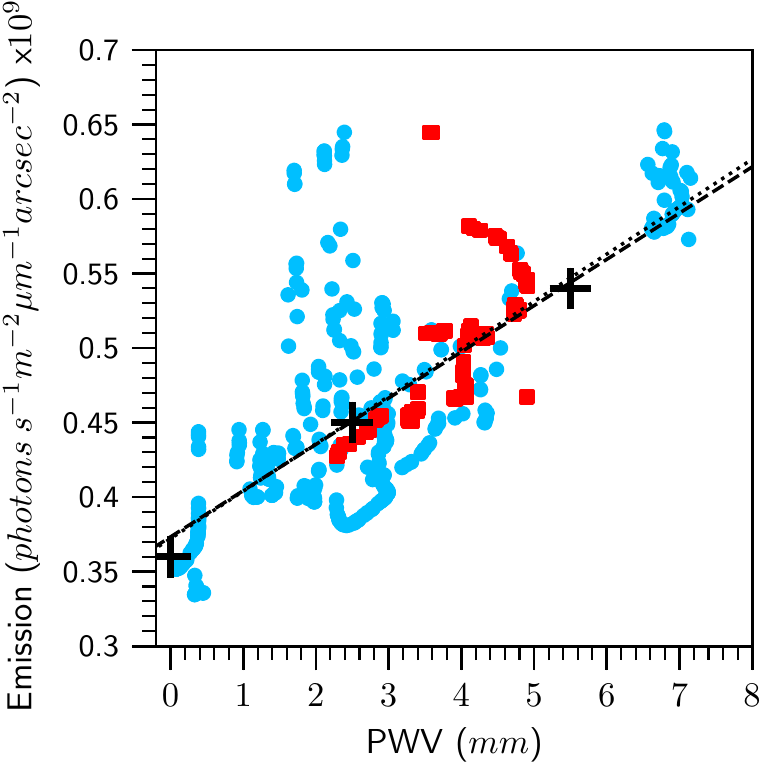}
\caption{Relation between emission from sky background obtained with VISIR instrument in the NEAR experiment and the observed PWV from LHATPRO. Blue dots represent the data obtained with the commissioning data-set (419 points), red squares are from the SV data-set (101 points). The dashed line is the linear regression computed only for the commissioning data. The dotted line is computed for all the data from commissioning and SV data-set. The crosses represents the values obtained with ESO's SKYCALC tool.}
\label{fig:backgr}
\end{figure}

We first analysed the data-set related to the commissioning period to compute the correlation between the sky background counts and the PWV.
Observations have been done with a angular groove phase mask coronograph (AGPM) optimised for the N band. The AGPM is a variation of a vortex coronagraph with very small inner working angle and high throughput \cite{mawet2005}. We considered the following technical specifications: an exposure time DIT = 6 msec, a $\Delta$$\lambda$ = 2.5 $\mu$m, a pixelscale = 0.045\arcsec, an efficiency = 0.1 (including the instrument transmission and the telescope reflectivity) and a GAIN = 20. To pass from counts to physical quantities we took into account the bias of the system of 10 kcts. This led to Fig.\ref{fig:backgr}, where blue dots represent the scattering plot between the sky-background and the observed PWV. We note that, as expected, the sky-background in the [10 - 12.5] $\mu$m window is coherent with that shown in Kasper et al. (2017) - Fig.4 in terms of order of magnitude of the emission mainly below 0.5-0.6 $\times$10$^{9}$ phot s$^{-1}$ m$^{-2}$ $\mu$m$^{-1}$ arcsec$^{-2}$ .

The interesting results is however that it is possible to observe a general correlation between the sky-background and the PWV which seems to be valid on a wide range of PWV values from 0 up to 7 mm. The spikes around PWV = 2 mm might be explained by emission due to some molecular absorption lines or interaction with dust in the atmosphere as described in \cite{kasper2017} - Fig.4. The contribution to these few peaks around PWV = 2 mm comes from different observations on different nights, so it can't be related to an isolated case and is therefore probably realistic.


The evidence of a sky background dependency on the PWV is supported by the fact that the cold Lyot stop blocks emission from the room temperature telescope environment (central obscuration, M2 support structures, etc). This only leaves blackbody radiation from the warm telescope mirrors and the cryostat entrance window, which should be comparable to the sky emission (see \citet{kasper2017}). The point in which the dashed black line intersects the Y-axis gives an indication of the telescope emission attenuated by the Lyot stop. The value of intersection with the Y-axis is of the order of 0.37$\times$10$^{9}$ that is somehow lower than estimated by \citet{kasper2017} ($\sim$0.6$\times$10$^{9}$). 

After the first analysis was completed, we considered also the data-set of the science verification (SV) to enrich the statistics, taking care to normalise with respect to the same observational conditions. During the SV a different Lyot stop has been used with respect to the commissioning run (NEAR-201 vs NEAR-101) permitting a larger throughput for signal and sky-background (93 per cent instead of 75 per cent). Emission has therefore to be corrected by a 75/93 factor to be compared with the commissioning data set. The red squares in Fig. \ref{fig:backgr} show results obtained with the SV data-set. We can see that results with bot data-sets are quite consistent, the  general correlation between the emission and the PWV is confirmed and the regression line is only slightly changed by the addition of SV data. Specifically the coefficients of the $y=ax+b$ regression line change from $[a=3.10\times10^7$, $b=3.73\times10^8]$, obtained with the commissioning data only, and $[a=3.18\times10^7$, $b=3.72\times10^8]$, obtained on the whole dataset. We also verified that the NEAR data is consistent with the sky background increase with PWV predicted by ESO's SKYCALC tool\footnote{\url{http://www.eso.org/sci/software/pipelines//skytools/skycalc}}. The crosses in Fig. \ref{fig:backgr} show the predicted values assuming the total mirror and warm side of the cryostat emissivity of 7\% and an instrument transmission of 50\%.\\

The important conclusion is therefore that, the knowledge in advance of the PWV can provide an indirect information of the sky-background regimes, allowing the identification of the conditions with the optimal sky background for the most challenging observations. As we have shown in this paper, the PVW can be quantified in a very accurate way on a time scale of few hours, thus our results indicate that it is indeed possible to predict the order of magnitude of the sky-background.

At the same time, this analysis puts in evidence that the correlation between the sky background and the PVW seems to be an important element to take into account in the design and realisation of METIS\footnote{METIS=Mid Infrared ELT Imager and Spectrograph}, a prime focus instrument conceived for the European Extremely Large Telescope (E-ELT) \citep{stuik2016}.

\section{Conclusions}
\label{concl}

In this study we aim to improve PWV forecasts at short time scales using a method recently proposed for this goal and to be applied at atmospheric and astroclimatic parameters \citep{masciadri2020}. The goal of the study is to analyse the impact of this strategy on the PWV at Cerro Paranal, site of the VLT, and in particular the implications in terms of science operations. This method is based on an auto-regressive technique. It makes use of the forecasts extended on a long time scale, performed in our case with a mesoscale model (we call this 'standard forecast') and computed in advance, and the real-time measurements obtained by the telescope instrumentation. In this paper we have shown that this method significantly increases the accuracy of the PWV forecasts by a factor that can be as high as $\sim$8.38 on a time scale of 1h. This means a gain of $\sim$800 per cent. Results obtained above Mt.Graham analysing other atmospheric parameters and seeing revealed consistent gains but not as high as the one obtained for the PWV in this study.

More precisely, we proved that, in the extremely low PWV regimes (PWV $\leq$ 1~mm) which are the most interesting for ground based IR astronomy, the forecast accuracy of our method achieves a negligible error (RMSE $\sim$ 0.04~mm) with respect to the measurements, for forecasts at a time scale of 1 hour. The RMSE is only slightly larger (The gain decreases in intensity if we consider larger time scales (2h, 3h, etc) but it remains still advantageous with respect to the standard forecast up to around 6 hours in the future. We also proved that the method revealed to be much more performant than the forecast by persistence that takes into account only real-time measurements. Accuracies are still extremely good for different PWV categories. For PWV $\leq$ 15 mm we obtain a RMSE = 0.13 mm  For PWV $\leq$ 5 mm we obtain a RMSE = 0.08 mm. By knowing that the accuracy of the measurements is 0.1 mm we can conclude that we basically achieved at least the limit of the instrumental accuracy with forecasts at 1h time scale for all the cases: PWV $\leq$ 15~mm,  PWV $\leq$ 5~mm and PWV $\leq$ 1~mm.

The standard forecast is still essential to long-term planning. Also it plays of course a role in the accuracy of the AR method. We implemented such a method in an automatic tool in order to assess the reliability for a potential operational support of VLT observations\footnote{At this stage this is not an official ESO forecast system.}. Results obtained in this study clearly indicate that such a kind of system can definitely play an important role in VLT observations scheduling and improve scientific throughput of critical IR observational campaigns. 

Besides that, we tested our system during the NEAR project commissioning and Science Verification and we discussed a potential further application of the PWV forecast. This has been done using NEAR project commissioning and Science Verification data-set related to a total sample of 25 nights. We found that the PWV values are directly correlated to the sky background IR emission in the [10-12.5] $\mu$m wavelength window in which the NEAR project runs, which is the optimal window in which molecular absorption from other atmospheric constituents is minimal. This evidence tells us that the knowledge in advance of  the PWV can be considered an indirect estimate of the sky background and this might open to further applications of the PWV forecasts. Of course it should be nice to confirm these results with a richer statistical sample.

The ability to accurately predict the PWV, and thus to have an indirect prediction of the sky background emission, could have a positive impact on IR observations and on the exploitation of the next-generation ELT IR instruments such as METIS.

\section*{Acknowledgements}

The authors thank the NEAR experiment team for giving access to the sky background measurements. The authors also thanks Filippo Mannucci for the useful discussions. This research has received funding from the European Union's Horizon 2020 research and innovation programme under grant agreement No 824135 (SOLARNET). Initialisation data come from the HRES atmospheric general forecast model of the ECMWF.

\section*{Data Availability}
Model initialisation data come from ECMWF atmospherical general forecast model HRES\footnote{\url{https://confluence.ecmwf.int/display/FUG/12.D++Model+Information}}. All NEAR project data are publicly available without restriction from archive.eso.org\footnote{\url{http://archive.eso.org/eso/eso_archive_main.html}}. The data produced by the model simulations can be shared on reasonable request to the corresponding author.








\appendix
\section{Cerro Paranal climatology}
\label{climatology}
\begin{figure}
\centering
\includegraphics[width=0.5\textwidth]{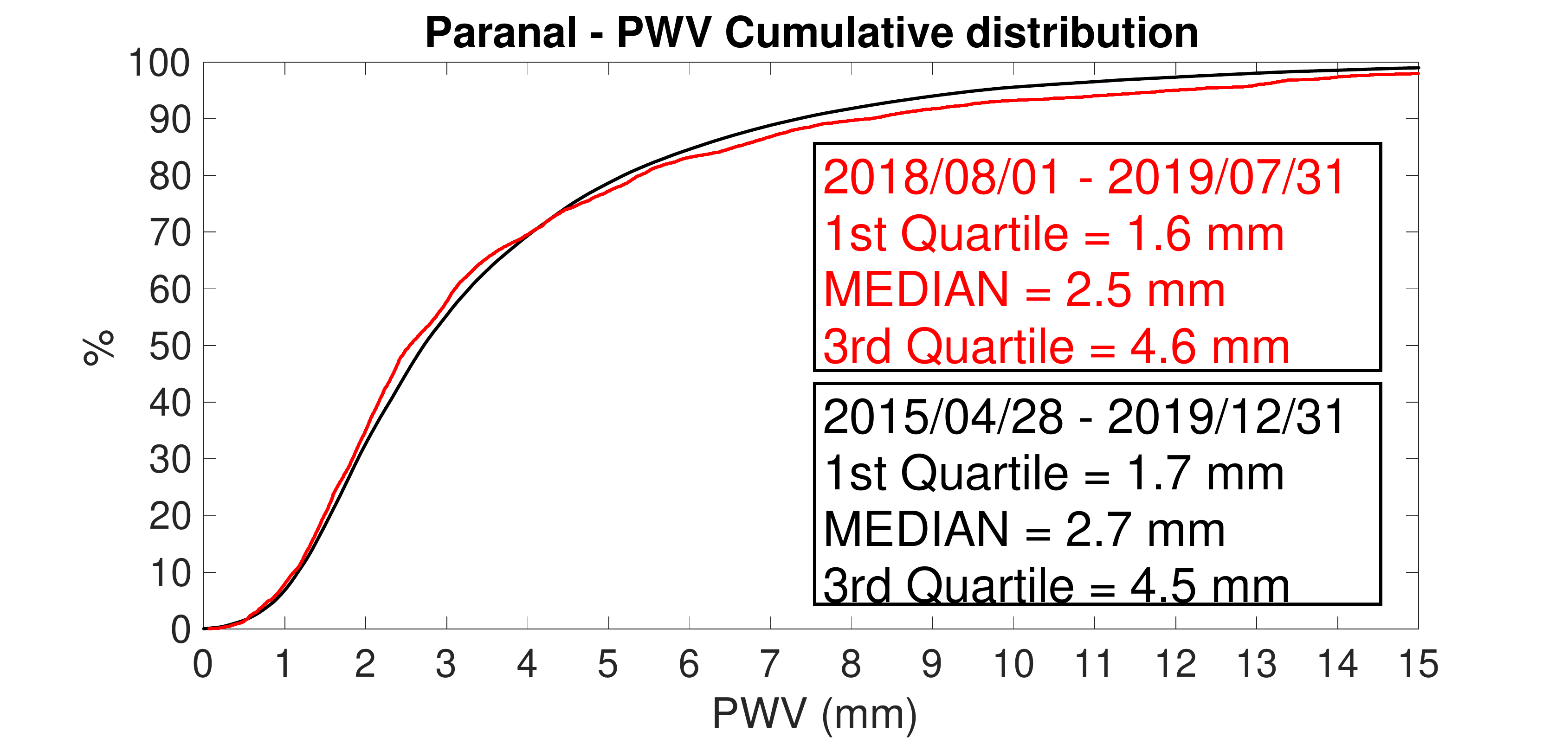}
\caption{Cumulative distribution of PWV over the 355 nights sample from 2018/2019 (red) used in this study, and on the 5 years 2015-2019 (black), as measured by LHATPRO}
\label{fig:cumdist}
\end{figure}

\begin{figure}
\centering
\includegraphics[width=0.5\textwidth]{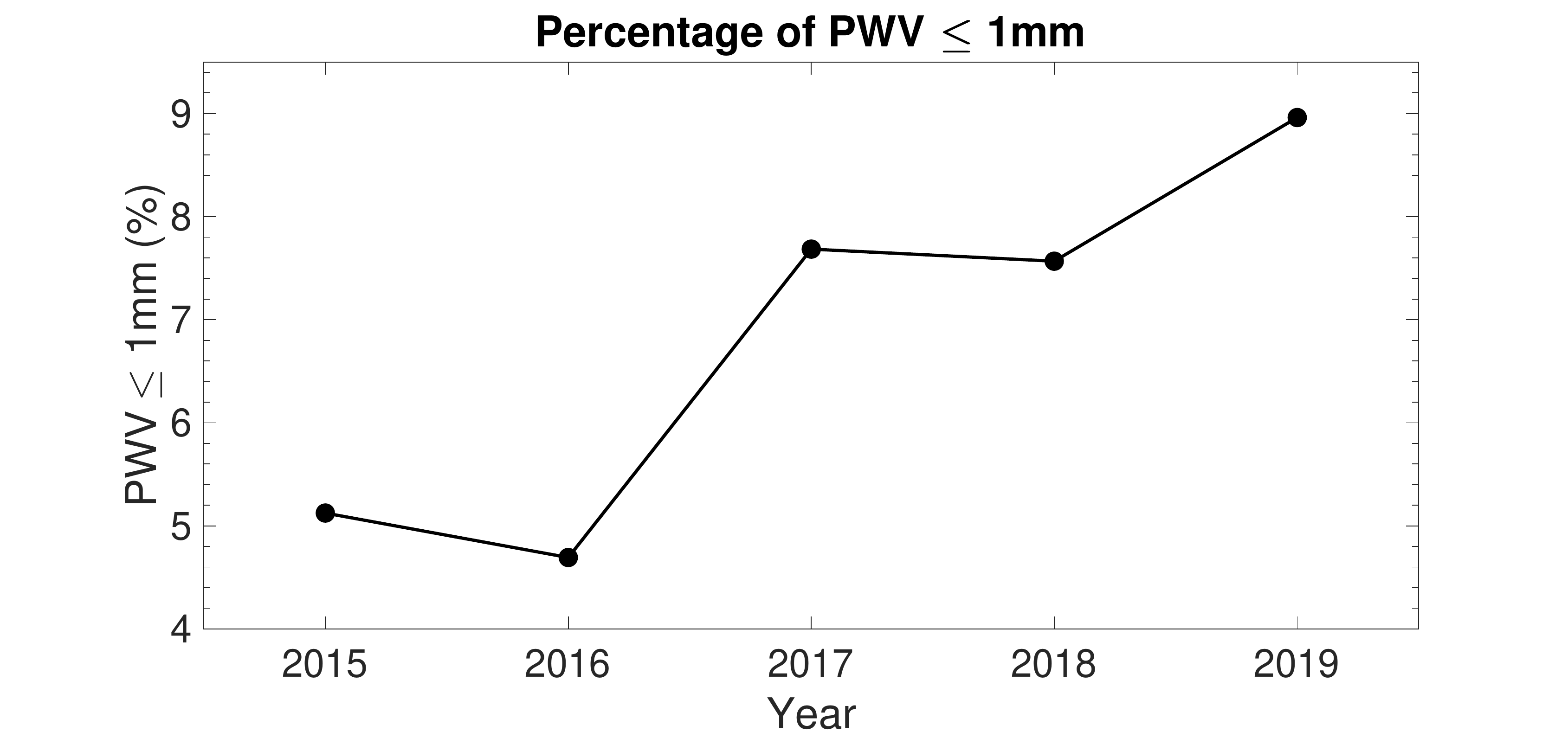}
\caption{Evolution of the percentage of time with PWV$\leq 1$~mm on Cerro Paranal between 2015 and 2019, with respect to the total PWV observations in each year.}
\label{fig:pevol}
\end{figure}

We report here the cumulative distributions of PWV obtained on the 355 nights sample from 2018/2019 used in the present study (fig. \ref{fig:cumdist}, red line). This is coherent with the statistics published in \citep{turchi2019} that treated data in the years 2013 and 2017. For the sake of completeness we also report the cumulative distribution of the PWV values on Cerro Paranal from 2015/04/28, which is the earliest LHATPRO measurement available in ESO database, to 2019/12/31 (fig. \ref{fig:cumdist}, black line). We observe that the chosen sample for this study is quite representative of the whole global sample. In fig. \ref{fig:pevol} we report the evolution of the percentage of time at Paranal characterised by extremely low PWV values $\leq 1$~mm, with respect to the total PWV observations in each year between 2015 and 2019. While the number of years monitored by LHATPRO is still low for a proper climatological analysis, it is enough to conclude that we can expect a percentage of extremely low PWV events in the range 5-10\% of the total time each year.

\bsp	
\label{lastpage}
\end{document}